\documentclass[letterpaper,11pt,twocolumn,oneside]{article}
\usepackage[dvips]{color}
\usepackage{graphicx}
\usepackage{amsbsy}
\usepackage[scanall]{psfrag}
\newcommand{\mnt}{\mbox{$m_{nt}$}}

\newcommand{\mntavg}{\mbox{$\langle m_{nt} \rangle$}}
\newcommand{\ddg}{\mbox{$\Delta\Delta G$}}
\newcommand{\ddga}{\mbox{$\Delta\Delta G^{a}$}}
\newcommand{\ddgb}{\mbox{$\Delta\Delta G^{b}$}}
\newcommand{\ddgm}{\mbox{$\Delta\Delta G^{m}$}}
\newcommand{\dddgm}{\mbox{d$\left(\Delta\Delta G^{m}\right)$}}
\newcommand{\dgf}{\mbox{$\Delta G_{f}$}}
\newcommand{\dgfmut}{\mbox{$\Delta G_{f}^{\rm{mut}}$}}
\newcommand{\dgfwt}{\mbox{$\Delta G_{f}^{\rm{wt}}$}}
\newcommand{\dgfextra}{\mbox{$\Delta G_{f}^{\rm{extra}}$}}
\newcommand{\conf}{\mbox{$\mathcal{C}$}}
\newcommand{\conft}{\mbox{$\mathcal{C}_t$}}
\newcommand{\ra}{\mbox{$\rightarrow$}}
\newcommand{\nuaa}{\mbox{$\langle \nu_{\rm{aa}} \rangle$}}
\newcommand{\nunt}{\mbox{$\langle \nu_{\rm{nt}} \rangle$}}

\begin{document}

\begin{titlepage}
\renewcommand{\baselinestretch}{1.0}
\center
\LARGE

{\bf Thermodynamic Prediction of Protein Neutrality}
\bigskip

\Large

Jesse D. Bloom,\footnotemark[1]\ \footnotemark[2] \quad
Jonathan J. Silberg,\footnotemark[3] \quad
Claus O. Wilke,\footnotemark[2]\ \footnotemark[4] \quad \\
D. Allan Drummond,\footnotemark[2]\ \footnotemark[5] \quad 
Christoph Adami,\footnotemark[2]\ \footnotemark[4] \quad
and Frances H. Arnold \footnotemark[1]
\bigskip

\large
\footnotemark[1] Division of Chemistry and Chemical Engineering 210-41, \\
\footnotemark[2] Digital Life Laboratory 136-93, \\
\footnotemark[5] Computation and Neural Systems, \\
California Institute of Technology, \\ Pasadena, Californa 91125 \\
\bigskip 

\footnotemark[3] 
Department of Biochemistry and Cell Biology \\
Rice University \\
Houston, TX 77005 \\
\bigskip

\footnotemark[4] Keck Graduate Institute, \\ 535 Watson Drive, \\
Claremont, California 91711 
\bigskip

Corresponding authors: \\
Jesse Bloom: bloom@caltech.edu \\
Frances Arnold: frances@cheme.caltech.edu \\
California Institute of Technology \\
Mail Code 210-41 \\
Pasadena, CA  91125 \\
Fax: 626-568-8743 \\ 
\bigskip

In press in \textit{Proc. Natl. Acad. Sci.}
\\
Classification: Biological Sciences, Biochemistry
\\
Total pages including this one: 14
\\
Total number tables: 3
\\ 
Total number of figures: 4
\\
Total number of Supporting Information tables: 1
\\
Total number of Supporting Information figures: 1
\\
Words in abstract: 147
\\
Characters in paper: 43742

\end{titlepage}

\begin{abstract}
We present a simple theory which uses thermodynamic parameters to
predict the probability that a protein retains the wildtype structure
after one or more random amino acid substitutions.
Our theory predicts that for large numbers of substitutions
the probability that a protein retains its structure will decline exponentially
with the number of substitutions, with the severity of this decline determined
by properties of the structure.  Our theory also predicts that a protein
can gain extra robustness to the first few substitutions by increasing its
thermodynamic stability.  We validate our theory with simulations on 
lattice protein models and by showing that it quantitatively predicts
previously published experimental measurements on subtilisin and our
own measurements on variants of TEM1 $\beta$-lactamase.
Our work unifies observations about the clustering of functional
proteins in sequence space, and provides a basis for interpreting 
the response of proteins to substitutions in protein 
engineering applications.
\end{abstract}

\subsection*{Introduction}
The ability to predict a protein's tolerance to amino acid
substitutions is of fundamental importance in understanding 
natural protein evolution,  
developing protein engineering strategies, and 
understanding the basis of genetic diseases.  
Computational and experimental studies have 
demonstrated that both protein stability and structure affect a 
protein's tolerance to substitutions.
Simulations have 
shown that more stable proteins have a higher fraction of 
folded mutants~\cite{Broglia1999,Bornberg-Bauer1999,Wingreen2004,Xia2004b}
and that some structures are encoded by more sequences than 
others~\cite{Li1996,Shakhnovich1991,Govindarajan1996}.
Experiments have demonstrated that proteins can be extremely tolerant
to single substitutions; for example, 84\% of single-residue mutants of 
T4 lysozyme~\cite{Rennell1991} and 65\% of single-residue mutants
of \textit{lac} repressor~\cite{Markiewicz1994} were scored as 
functional.  For multiple substitutions, 
the fraction of functional proteins decreases roughly exponentially
with the number of substitutions, although the severity of this decline
varies
among proteins~\cite{Shafikhani1997,Guo2004,Daugherty1999}.
Protein mutagenesis experiments have also underscored the contribution of
protein stability to mutational tolerance by 
finding ``global suppressor'' substitutions that
buffer a protein against otherwise deleterious substitutions
by increasing its stability~\cite{Shortle1985,Poteete1997}.

We unify these diverse experimental and computational results 
into a simple framework for predicting
a protein's tolerance to substitutions.  
A fundamental measure of this tolerance is the fraction of proteins 
retaining the wildtype structure after a single random substitution, often
called the neutrality~\cite{Wilke2002}.  
We extend this concept to multiple substitutions by defining
the $m$-neutrality as the fraction of proteins that fold to the wildtype
structure among all
sequences that differ from the wildtype sequence at $m$ residues.
Since mutants that fail to fold also generally fail to function, the 
$m$-neutrality provides an upper bound to the fraction of proteins
with $m$ substitutions that retain biochemical function.
We show that a protein's $m$-neutrality can be accurately
predicted from measurable thermodynamic parameters, and that
these predictions capture the contributions of both
stability and structure to determining a protein's tolerance to substitutions.

\small
\subsection*{Methods}

\subsubsection*{Lattice Protein Model}
We performed simulations with lattice
proteins~\cite{Chan2002} of length $L = 20$ monomers of 
$20$ types corresponding to the natural amino acids.
We folded the proteins on a two-dimensional lattice, allowing them
to occupy any of the 41,889,578 possible compact or non-compact 
conformations.  The energy of a conformation 
\conf\ is the sum of the nonbonded nearest-neighbor interactions,
$$\label{eq:econf}
E\left(\conf\right) = \sum\limits_{i=1}^{L}\sum\limits_{j=1}^{i-2} C_{ij}\left(\conf\right) \times \epsilon\left(\mathcal{A}_i, \mathcal{A}_j\right),$$
where $C_{ij}\left(\conf\right)$ is one if 
residues $i$ and $j$ are nearest
neighbors in conformation \conf\ and zero otherwise, and 
$\epsilon\left(\mathcal{A}_i, \mathcal{A}_j\right)$ is the interaction
energy between residue types $\mathcal{A}_i$ and $\mathcal{A}_j$,
given by Table 5 of \cite{Miyazawa1985}.  

The primary advantage of using lattice proteins is that we
can exactly compute the stability of a conformation \conft\ as 
$$\dgf\left(\conft\right) = E\left(\conft\right) + T \ln \left\{Q\left(T\right) - \exp\left[-E\left(\conft\right)/T\right]\right\},$$
where $Q\left(T\right)$ is the partition sum 
$$Q\left(T\right) = \sum\limits_{\left\{\conf_i\right\}} \exp \left[-E\left(\conf_i\right)/T\right]$$
over all conformations, made tractable by noting that there are 
only 910,972 unique contact sets.
All simulations were performed
at a reduced temperature of $T = 1.0$

\subsubsection*{TEM1 $\beta$-Lactamase Mutant Libraries}
To examine the effects of mutations on the retention of protein
function, we constructed mutant libraries of wildtype and the 
thermostable M182T variant of TEM1 $\beta$-lactamase.
The 861 bp genes (a 
kind gift from Brian Shoichet~\cite{Wang2002}) were subcloned
into the pMON:1A2 plasmid~\cite{Meyer2003} 
with \textit{SacI} and \textit{HindIII} using PCR primers 
{\footnotesize 5'-GCGGCG\underline{GAGCTC}ATGAGTATTCAACATTTCCGT\\GTCGC-3'}
and {\footnotesize 5'-GCGGCG\underline{AAGCTT}TTACCAATG\\CTTAATCAGTGAGGCAC-3'}
(restriction sites are underlined).  
We first created a control unmutated library by cutting
the gene directly from the plasmid.  This unmutated gene was used as
the template for a round of error-prone PCR with 100 $\mu$l reactions
containing 3 ng of template, 
0.5 $\mu$M of each of the above primers, 7 mM MgCl$_2$, 75 $\mu$M
MnCl$_2$, 200 $\mu$M of dATP and dGTP, 500 $\mu$M
of dTTP and dCTP, 1X Applied Biosystems PCR buffer without MgCl$_2$, and
5 U of Applied Biosystems \textit{Taq} DNA polymerase.
The PCR conditions were 95$^o$C for 5 minutes, and then 14 cycles of
30 s each at 95$^o$C, 50$^o$C, and 72$^o$C.  The product from this PCR
was digested with \textit{SacI}/\textit{HindIII} and gel purified, and then used as the template
for another identical round of error--prone PCR.  This process was repeated
to create five libraries with increasing numbers of mutations, which we
labeled EP-0 (for the unmutated control) to EP-5 (for the 
product of the fifth round of error--prone PCR).
We quantified the number of doublings for each 
round by running PCR product versus a known standard on an agarose 
gel, and found that our protocol consistently yielded 
ten doublings.

To measure the fraction of genes in the mutant libraries that still 
encoded functional proteins, we ligated the 
genes into the pMON:1A2 plasmid
with T4 Quick DNA Ligase in 20 $\mu$l reactions containing 50 ng each
of gene and plasmid, and then transformed 5 $\mu$l of the ligation reactions 
into 50 $\mu$l of XL1-Blue Supercompetent cells from Stratagene.
The transformed
cells were plated on LB-agar plates containing 10 $\mu$g/ml of kanamycin
(selective only for plasmid) and on LB-agar plates containing
10 $\mu$g/ml of kanamycin and 
20 $\mu$g/ml of ampicillin (selective
for both plasmid and active TEM1 gene) at a density that
gave 100-300 colonies per unselected plate.  The fractions functional 
were computed as the average of at least five pairs of selected/unselected
plates, and are shown in Table 2.

\begin{table}
\centerline{\bf TEM1 mutation frequencies.} 
\medskip

\centerline{\begin{tabular}{|l|c|} \hline
Base pairs sequenced & 22,800 \\ \hline
Total mutations & 172 \\ \hline
Total AA substitutions & 120 \\ \hline
Mutation frequency (\%) & 0.75 $\pm$ 0.06 \\  \hline
Mutations per gene & 6.5 $\pm$ 0.5 \\ \hline
AA substitutions per gene & 4.5 $\pm$ 0.4 \\ \hline
Mutation types (\%) & \\ 
\qquad A \ra T, T \ra A & 22 \\ 
\qquad A \ra C, T \ra G & 9 \\ 
\qquad A \ra G, T \ra C & 42 \\ 
\qquad G \ra A, C \ra T & 20 \\ 
\qquad G \ra C, C \ra G & 1 \\ 
\qquad G \ra T, C \ra A & 3 \\ 
\qquad frameshift & 3 \\ \hline
\end{tabular}}
\caption{\label{tab:mutations}
Mutation frequencies for TEM1 $\beta$-lactamase mutagenesis
determined by sequencing
20 unselected clones each from the round five wildtype and 
M182T error-prone PCR 
libraries.}
\end{table}

To test the ability of our theory to predict the decline in $m$-neutrality,

The mutation frequency in the round five library was determined
by sequencing the first 570 bp of
twenty genes each from the unselected wildtype and M182T
plates with the sequencing 
primer {\footnotesize 5'-GGTCGATGTTTGATGTTATGGAGC-3'}.  
The wildtype and M182T genes were mutated under identical conditions,
and the sequencing found the same nucleotide mutation frequencies
for both (0.77 $\pm$ 0.08\% for wildtype and 0.74 $\pm$ 0.08\% for M18T2, 
corresponding to 6.6 $\pm$ 0.7 and 6.4 $\pm$ 0.7 
nucleotide mutations per 861 bp gene).
For better statistics, the sequencing results
for both libraries were combined to give the data in 
Table 1.  No biases in the locations of the mutations were observed.
Eleven mutations occurred twice, which is in good agreement with the
expectation of eight duplicate mutations if all possible mutations were
equiprobable.  The per-round mutation frequency was
calculated as 0.15 $\pm$ 0.03\% (1.3 $\pm$ 0.3 nucleotide mutations per gene) by assuming that each round of error-prone PCR introduced
the same average number of mutations.  To confirm this assumption,
we sequenced ten unselected clones each from the wildtype and M182T 
round one libraries,
and found mutation frequencies of 0.16 $\pm$ 0.05\% for wildtype
and 0.19 $\pm$ 0.06\% for M182T.
Standard errors were computed assuming Poisson sampling statistics.
More detailed sequencing
information is in Table 4 of the Supporting Information.

\normalsize
\subsection*{Results}

\subsubsection*{Thermodynamic Framework for Predicting Neutrality}
A protein's native structure is thermodynamically 
stable~\cite{Anfinsen1973,Ellis1998}, with typical free energies of
folding (\dgf) between $-5$ and $-15$ kcal/mol~\cite{Fersht1999}.
A mutant sequence folds to the wildtype structure only if the stability
of that structure meets some minimal threshold.
We call the extra stability of the native structure 
beyond this minimal threshold \dgfextra\ and note that
functional proteins always have $\dgfextra \le 0$. 
We define protein's $m$-neutrality as the fraction of
sequences with $m$ substitutions that still meet the stability threshold. 

A substitution causes a stability change of
$$\ddg = \dgfmut - \dgfwt$$
where \dgfwt\ and \dgfmut\ are the wildtype and mutant protein
stabilities.  Substitutions tend to be 
destabilizing: although there are no large collections of \ddg\ measurements 
for truly random substitutions, in a likely-biased collection of more than 
2,000 measured \ddg\ values for single-residue 
substitutions~\cite{Bava2004}, the mean is $0.9$ kcal/mol
and the values at the 10th and 90th percentiles 
are $-1.0$ and $3.2$.

The thermodynamic effects of most substitutions are approximately 
additive~\cite{Wells1990,Zhang1995,Serrano1993}, meaning that
if the stability changes due to two different 
single substitutions are \ddga\ and \ddgb, then
the stability change due to both substitutions is approximately
$\ddga + \ddgb$.
If we know the probability distribution $p_1\left(\ddg\right)$ that
a single random substitution causes a stability change of \ddg,
and if we assume that substitutions are additive, then the net effect
\ddgm\ of $m$ random substitutions
is just the sum of $m$ random variables from the probability distribution
$p_1\left(\ddg\right)$.  Under this additivity assumption, we can 
therefore directly calculate the 
distribution $p_m\left(\ddgm\right)$
for \ddgm\ by
performing an $m$-fold convolution~\cite{vanKampan1992} of $p_1\left(\ddg\right)$. 

The $m$-neutrality $P_f\left(m\right)$ is simply the
the probability that \ddgm\ is not more destabilizing
than the extra stability \dgfextra\ of the wildtype sequence, and
can be written as 
\begin{equation}
\label{eq:mneutrality}
P_f\left(m\right) = \int\limits_{-\infty}^{-\dgfextra} p_m\left(\ddgm\right) \dddgm.
\end{equation}
This formula gives  
a protein's $m$-neutrality in terms of its extra stability 
and the distribution of \ddg\ values for all possible single substitutions.

\subsubsection*{Lattice Proteins Support Predictions}
\begin{figure*}
\centerline{\includegraphics[width=6.5in]{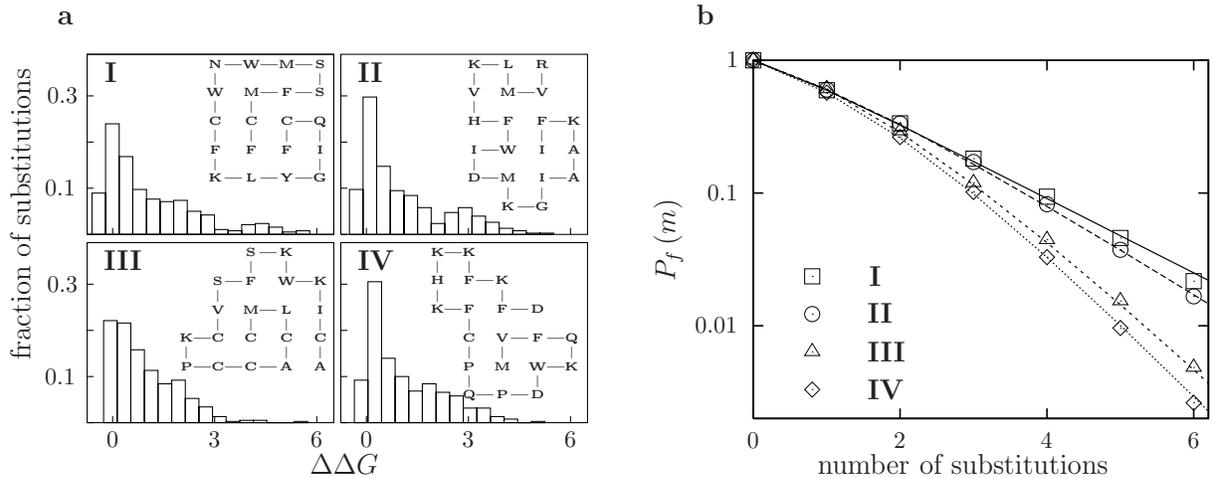}}
\caption{\label{fig:diffstructs}
Lattice proteins with
different structures but the same stability ($\dgf = -1.0$) converge
to different exponential declines in $m$-neutrality.
{\bf (a)} The distributions of \ddg\ for all $380$
single amino acid substitutions to the inset lattice proteins.
{\bf (b)} The measured (symbols) and predicted (lines)
$m$-neutralities for the four proteins.  Proteins
are considered folded if $\dgf \le 0.0$ for the original
native structure. The proteins used for the
$m$-neutrality analyses were generated by adaptive walks from 
random starting sequences, followed by $2.5 \times 10^5$
generations of neutral evolution with a population size
of 100 and a per generation per residue substitution rate of
$5 \times 10^{-5}$, selecting
for sequences with $\dgf \le -1.0$ and then taking the first
sequence generated with a stability within 0.025 of -1.0.
The $m$-neutralities were computed by sampling all mutants for
$m \le 2$ or $5 \times 10^5$ random mutants for $m > 2$.  The
predicted $m$-neutralities were computed according to Eq. 
\ref{eq:mneutrality}
by numerically convolving the distribution of single--substitution
\ddg\ values using generating functions~\cite{vanKampan1992} computed
with fast--Fourier transforms and a bin size of 0.01.}
\end{figure*}

We tested the ability of this simple 
framework to predict the fraction of lattice 
proteins that retained the original structure after random amino acid 
substitutions.  Lattice proteins are highly simplified models
of proteins that provide a useful tool for studying protein
folding~\cite{Dill1995,Hinds1994,Shakhnovich1993,Socci1998}
and evolution~\cite{Chan2002,Xia2004} (some example lattice proteins
are shown in Figure \ref{fig:diffstructs}).  We can easily measure the
$m$-neutralities of the lattice proteins by making random amino acid
substitutions and seeing if the sequences still have $\dgf \le 0.0$. 
We can also use Eq.
\ref{eq:mneutrality} to directly predict the $m$-neutralities
since we can exactly compute
\dgf\ and \ddg\ values.

Eq. \ref{eq:mneutrality} accurately predicted the
$m$-neutralities of all of the lattice proteins we tested.
Lattice proteins with different structures have different
$m$-neutralities, even when they have the same \dgf\ (Figure \ref{fig:diffstructs}).
The $1$-neutralities of proteins with different
structures and the same \dgf\ look similar, but for larger 
values of $m$ some proteins
clearly show higher $m$-neutralities than others.  For large 
$m$, the $m$-neutralities of all of the proteins converge to a simple
exponential of the form
$$P_f\left(m\right) \propto \nuaa^{m}$$
where \nuaa\ is the average fraction of proteins that are destabilized
by a further single random amino acid substitution after several 
substitutions have already occurred.
The underlying reason for the exponential form of this decline is clear:
after several substitutions the distribution of \dgf\ among
the remaining functional sequences reaches a steady state
and each new substitution pushes the same fraction of proteins
beyond the stability threshold.  The average neutrality \nuaa\ is
therefore actually the $1$-neutrality averaged over all stable 
sequences with the wildtype structure.
Although $P_f\left(m = 1\right)$ is similar for all
of the protein structures 
in Figure \ref{fig:diffstructs}, the factors that give
rise to the different values of \nuaa\ for the different structures 
are present 
in the distribution of single mutant \ddg\ values,
since it is used to predict the 
$m$-neutralities for all values of $m$.

\begin{figure}
\centerline{\includegraphics[width=3.1in]{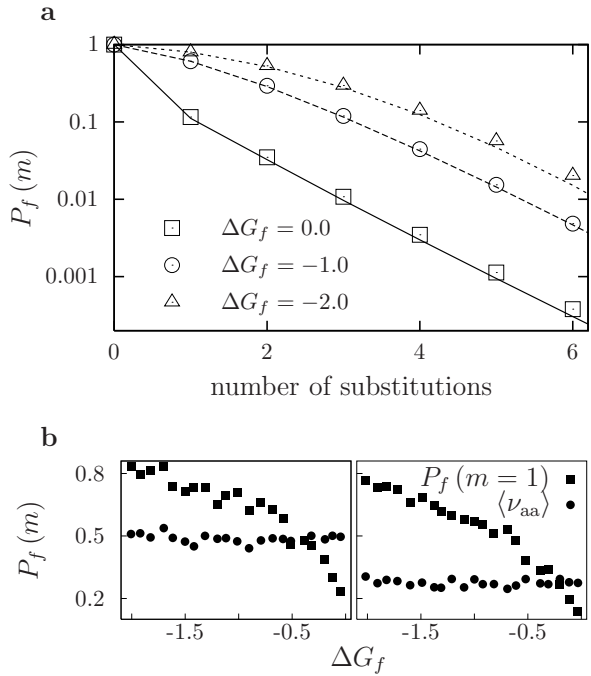}}
\caption{\label{fig:diffstabs}
Lattice proteins with the same structure but different stabilities have
different $1$-neutralities but have the same average neutrality \nuaa.
{\bf (a)} Predicted (lines) and measured (symbols) $m$-neutralities
for proteins with different stabilities and the same structure 
({\bf III} in Figure \ref{fig:diffstructs}).
{\bf (b)} Measured values of the $1$-neutralities (squares) 
and average neutralities
(circles) for proteins with different stabilities
but the same structures (the plots at left and right are for 
structures {\bf I} and {\bf IV} from Figure 
\ref{fig:diffstructs}, respectively).  The sequences were generated
by finding a sequence with $\dgf = -2.0$ using
the procedure described in Figure \ref{fig:diffstructs}, and then
using this sequence as a starting point for neutral evolution 
selecting for the indicated target stabilities.  The proteins
with different stabilities are highly diverged, with average pairwise 
sequence identities of 15\% and 41\% for the structures at left
and right, respectively.  The $m$-neutralities were computed
as in Figure \ref{fig:diffstructs}, and \nuaa\ was computed
as the square root of the $6$-neutrality divided by the $4$-neutrality.}
\end{figure}

Figure \ref{fig:diffstabs} shows the $m$-neutralities of proteins with
the same structure but different stabilities.  After 
several substitutions, all
of the proteins converge to  
the same value of \nuaa, suggesting that \nuaa\
is a generic property of a protein's structure.
On the other hand, the response of
a protein to the first few substitutions depends strongly on its stability,
with more stable proteins exhibiting higher initial $m$-neutrality. 
The high initial
$m$-neutrality of stable proteins is readily rationalized in terms of 
the thermodynamic model:
substitutions tend to disrupt a protein's structure
by pushing its stability below the minimal threshold, but proteins with 
an extra stability cushion are buffered against
the first few substitutions~\cite{Axe2004}.  Proteins that
sit on the very margin of the minimal stability threshold exhibit 
lower $1$-neutrality than is predicted by an exponential decline
because these proteins are less stable than the average folded
protein, and so surviving sequences
will tend to be more stable than the wildtype sequence
and so be more tolerant to the next substitution.

\begin{table*}
\centerline{\bf TEM1 mutant library measurements.}
\medskip
 
\centerline{\begin{tabular}{|c|c|c|c|c|}
\hline
{\bf Round} & $\pmb{\langle m_{nt} \rangle}$ & $\pmb{\langle m_{aa} \rangle}$ & {\bf WT} & \bf{M182T}\\ \hline 
0 & 0.0 $\pm$ 0.0 & 0.0 $\pm$ 0.0 & 0.76 $\pm$ 0.03 & 0.74 $\pm$ 0.04 \\ 
1 & 1.3 $\pm$ 0.2 & 0.9 $\pm$ 0.1 & 0.59 $\pm$ 0.03 & 0.68 $\pm$ 0.03 \\ 
2 & 2.6 $\pm$ 0.3 & 1.8 $\pm$ 0.2 & 0.47 $\pm$ 0.03 & 0.54 $\pm$ 0.02 \\ 
3 & 3.9 $\pm$ 0.4 & 2.7 $\pm$ 0.2 & 0.28 $\pm$ 0.02 & 0.45 $\pm$ 0.04 \\ 
4 & 5.2 $\pm$ 0.4 & 3.6 $\pm$ 0.3 & 0.18 $\pm$ 0.01 & 0.28 $\pm$ 0.01 \\ 
5 & 6.5 $\pm$ 0.5 & 4.5 $\pm$ 0.4 & 0.13 $\pm$ 0.01 & 0.20 $\pm$ 0.02 \\ \hline
\end{tabular}}
\caption{\label{tab:TEM1} Measured fractions of functional proteins
in mutant libraries
of wildtype and the thermostable M182T variant of TEM1 $\beta$-lactamase.
The table shows the number of rounds of error-prone
PCR, the average number of nucleotide mutations per gene, and the fractions
of mutated genes that confer ampicillin resistance in 
\textit{E. coli}.
Values are shown $\pm$ their standard
errors.}
\end{table*}

\subsubsection*{Real Proteins Support Predictions}
Our theory makes two main predictions:
first, that the decline in $m$-neutrality is determined by
the \ddg\ values for single amino acid
substitutions, and second, that among proteins with the same structure,
more stable variants will have higher $m$-neutralities.  We tested
these predictions against measurements of the fractions of functional
proteins in mutant libraries of subtilisin and variants of TEM1
$\beta$-lactamase.  Our theory
is designed to predict the fraction of proteins that retain the
wildtype structure, but the experiments measure the fraction of
proteins that retain function.  However, since proteins that
fail to fold also generally fail to function, our theory provides
an upper bound on the fraction of functional proteins.  We expect 
that for many proteins this upper bound will closely approximate the
actual fraction functional since mutagenesis studies suggest
that most functionally disruptive random substitutions disrupt
the structure rather than specifically affect functional
residues~\cite{Shortle1985,Loeb1989,Pakula1986}.

To test the ability of our theory to predict the decline in $m$-neutrality,
we used data on the fractions of functional proteins in subtilisin mutant 
libraries created by Shafikhani and coworkers~\cite{Shafikhani1997}
(population 6B of Table 2 of \cite{Shafikhani1997}, normalized by the fraction of functional
clones in the control libraries) and our own mutant libraries of
TEM1 (Table 2).  Each mutant library contains a distribution
of sequences with different numbers of amino acid mutations.  
The form of this distribution
is known: the probability that a sequence in a library
with an average of \mntavg\ nucleotide mutations created by $N$ cycles
of PCR with a PCR efficiency of $\lambda$ will have \mnt\ mutations is
$$f\left(\mnt\right) = \left(1 + \lambda\right)^{-N} \sum\limits_{k = 0}^{N} {N \choose k} \lambda^k \frac{\left(kx\right)^{\mnt} e^{-kx}}{\mnt!}$$
where $x = \mntavg \left(1 + \lambda\right) / \left(N \lambda\right)$
~\cite{Sun1995,Drummond2004}.  Subtilisin was mutagenized using 13 PCR
cycles with 10 effective doublings~\cite{Shafikhani1997}, so
$N$ is $13$ times the number of rounds of error--prone PCR
and $\lambda = 0.77$.
TEM1 was mutagenized using 14 PCR cycles with 10 effective doublings,
so $N$ is $14$ times the number of rounds and $\lambda = 0.71$.
We confirmed 
that $f\left(\mnt\right)$ accurately describes the distribution 
of mutations in our libraries (Figure 5 of the Supporting Information).

The expected fraction of folded sequences in a mutant library is
easily calculated from $f\left(\mnt\right)$ and the probability
$P_f\left(\mnt\right)$ that a sequence is still functional
after \mnt\ nucleotide mutations as
$$\mathcal{F} = \sum\limits_{\mnt = 0}^\infty f\left(\mnt\right) \times P_f\left(\mnt\right).$$

We calculated the probability $P_f\left(\mnt\right)$ that
a sequence was still folded after \mnt\ nucleotide mutations
by using two existing computer programs for estimating the
\ddg\ values for single substitutions to proteins with known 
structures (PDB structure 1IAV for subtilisin and 1BTL for TEM1):
Gilis and Rooman's PoPMuSiC potential~\cite{Gilis2000} and Serrano and
coworkers' FOLDEF potential~\cite{Guerois2002} with van der Waals clash
energies.  Since the genetic code makes nucleotide mutations more
likely to induce some amino acid substitutions than others, and since
error--prone PCR introduces a non-random distribution of nucleotide
mutations, we weighted each \ddg\ value by the probability that
it would be induced by a single nucleotide mutation made according
to the observed error--prone PCR nucleotide 
mutation frequencies (given in Table 1
of \cite{Shafikhani1997} for subtilisin and Table 1 of the current
work for TEM1).  We assigned a \ddg\ of zero to 
synonymous nucleotide mutations
since they do not cause an amino acid substitution,
and we assigned a \ddg\ of $25$ kcal/mol to frameshift and 
nonsense mutations since premature truncation is
expected to inactivate the protein.  We ignored the small number
of substitutions for which PoPMuSiC failed to calculated a \ddg.
With this weighted
\ddg\ distribution for nucleotide mutations, all we needed to construct
$P_f\left(\mnt\right)$ according to Eq. \ref{eq:mneutrality}
was the value of \dgfextra.  This cannot be measured directly since
we do not know the minimal stability threshold.  
However, since \dgfextra\ only
influences the initial behavior of the $m$-neutrality and
does not affect the limiting decline (Figure 2), and since we
have six data points for each protein, we could do a least-squares fit 
of \dgfextra\ to
the data and still test the ability of the theory to predict
the decline in the fraction of functional proteins. 

\begin{figure}
\centerline{\includegraphics[width=3.1in]{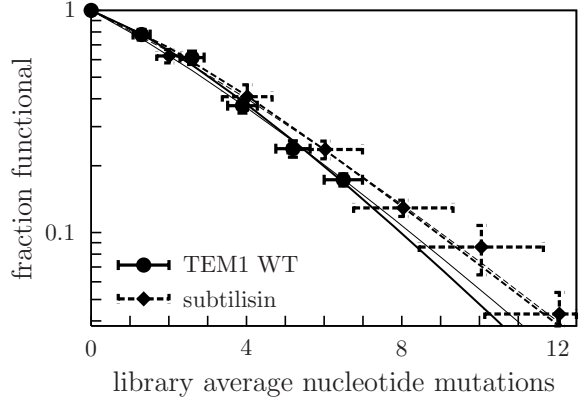}}
\caption{\label{fig:realproteins} Theoretical predictions and fractions of
functional proteins in mutant libraries of subtilisin (dashed lines) and 
TEM1 $\beta$-lactamase (solid lines) genes.
Thick lines show predictions made using PoPMuSiC~\cite{Gilis2000}
and thin lines show predictions made using FOLDEF~\cite{Guerois2002}.
The TEM1 measurements are from Table 2, normalized by the
values from the control unmutated library.}
\end{figure}

Figure \ref{fig:realproteins} shows the measured fractions of functional
proteins for subtilisin and wildtype TEM1 versus
the theoretical predictions made with PoPMuSiC and FOLDEF. 
The theoretical predictions closely match the measured fractions
of functional proteins in all cases, with subtilisin exhibiting slightly
higher $m$-neutralities than TEM1.  

\begin{figure}
\centerline{\includegraphics[width=3.1in]{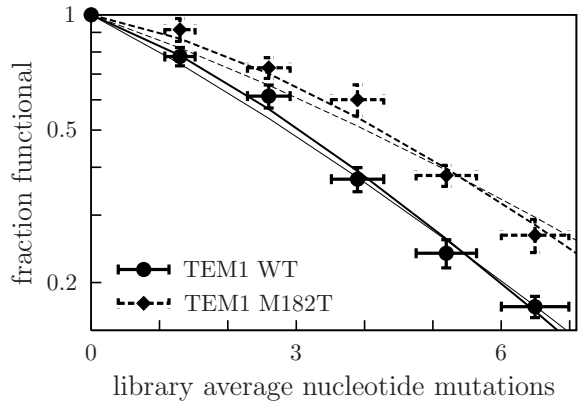}}
\caption{\label{fig:TEMvariants} The more stable M182T variant of
TEM1 $\beta$-lactamase (dashed lines) exhibits a higher fraction
of functional mutants relative to wildtype (solid lines), as
predicted.
Thick lines show predictions made using PoPMuSiC~\cite{Gilis2000}
and thin lines show predictions made using FOLDEF~\cite{Guerois2002}.
The measurements are from Table 2, normalized by the
values from the control unmutated library.}
\end{figure}

The second major prediction of our theory is that among proteins
with the same structure, more stable variants
will exhibit higher initial $m$-neutralities,
but converge to same average neutrality.  To test
this prediction, we compared the fractions of functional proteins
in mutant libraries of wildtype and the M182T variant of TEM1.
The M182T variant differs from wildtype by only a single substitution
yet is 2.7 kcal/mol more stable~\cite{Wang2002}, so we predict that it should
exhibit a higher fraction of functional proteins at the same level 
of mutation.
Figure \ref{fig:TEMvariants} shows the measured fractions functional
for wildtype and the M182T variant, as well as the theoretical predictions
made using both PoPMuSiC and FOLDEF.  
As predicted, the M182T variant exhibits a higher fraction of 
functional proteins, and once
again the predictions made with both potentials are in good agreement
with the experimental measurements.

\begin{table*}
\centerline{\bf Predicted Average Neutralities} 
\medskip

\centerline{\begin{tabular}{|l|l|l|c|c|c|} \hline
{\bf PDB} & {\bf Protein} & {\bf CATH architecture} & {\bf Length} & \pmb{\nunt} & \pmb{\nuaa} \\
\hline
1IAV & subtilisin & $\alpha\beta$ 3-layer sandwich & 269 & 0.65 & 0.55 \\
\hline
1B9C & GFP & $\beta$ barrel & 236 & 0.62 & 0.56 \\ 
\hline
1BTL & TEM1 $\beta$-lactamase & $\alpha\beta$ 3-layer sandwich & 263 & 0.58 & 0.46 \\
\hline
1RLV & tRNA endonuclease & not classified & 305 & 0.55 & 0.44 \\
\hline
1HZW & thymidylate synthase & $\alpha\beta$ 2-layer sandwich & 290 & 0.50 & 0.41 \\
\hline
2BNH & ribonuclease inhibitor & $\alpha\beta$ horseshoe & 457 & 0.45 & 0.35 \\
\hline
1HEL & hen lysozyme & $\alpha$ orthogonal bundle & 129 & 0.43 & 0.38 \\
\hline
\end{tabular}}
\caption{\label{tab:predictions}
Predictions of the average neutralities of various proteins to 
both nucleotide mutations (\nunt) and amino acid substitutions (\nuaa).  
The codes for the PDB structures and the CATH~\cite{Orengo1997}
architectures are shown along with the lengths of the protein
chains in the PDB structures (in all cases we consider chain A).
The average neutralities are computed by calculating
the fraction of sequences predicted be inactivated by the 
$10$th mutation or substitution, using \ddg\ values from
PoPMuSiC~\cite{Gilis2000} and 
assuming that the proteins all have the same value 
of \dgfextra\ as wildtype TEM1 $\beta$-lactamase.  The
values of \nunt\ are computed assuming that nucleotide mutations
are made according to the 
error--prone PCR mutation frequencies of Table 1.}
\end{table*}

To further explore the range of possible neutralities for different
proteins, we used \ddg\ values from PoPMuSiC to predict
the expected average neutralities to both amino acid
substitutions (\nuaa) and nucleotide mutations (\nunt)
for proteins chosen from several
different CATH~\cite{Orengo1997} protein structure classifications.
Since we do not know \dgfextra\ for these proteins, we computed the
fraction of proteins expected to be inactivated by the $10$th mutation 
since after this many mutations 
effects due to the initial protein stability should be small.
Table 3 shows the predicted average neutralities to both random amino acid
substitutions and nucleotide mutations made according to the
mutation probabilities of our TEM1 mutagenesis.  The predicted average
neutralities
differ considerably, showing that our theory predicts that different
proteins can have substantially different neutralities.

\subsection*{Discussion}
We have presented a theory for calculating the probability that a
protein will retain its structure after random 
amino acid substitutions,
and have confirmed the main theoretical
predictions with simulations and experiments.  Our theory 
naturally separates a protein's $m$-neutrality 
into components due to structure
and stability.  The eventual severity of the exponential decline in 
$m$-neutrality with the number of substitutions is
a property of a protein's structure. 
On the other hand, increased stability confers greater tolerance to the first 
few substitutions, in effect
allowing a protein to ``take a few hits'' before it is 
pushed into the inevitable structurally determined
exponential decline in $m$-neutrality.  This increased
tolerance to mutations due to extra stability is probably
also the underlying reason for the existence of global suppressor
mutations~\cite{Shortle1985,Poteete1997} 
that buffer proteins against otherwise deleterious
mutations.

The major assumption underlying our theory is that the thermodynamic
effects of substitutions are additive.  This assumption
is clearly not strictly true since protein residues do interact.
Substitutions are most likely to be non-additive
if the mutated residues are in close contact
in a protein's structure~\cite{Wells1990,Zhang1995}.   
Since proteins are large, two randomly chosen residues will
rarely contact each other, and so although the additivity assumption
is certainly violated for some specific combinations 
of substitutions, it is
accurate when averaged over all possible substitutions.
When we apply our theory to measurements of the fraction of 
mutant proteins that retain function we are making a second assumption
by ignoring the possibility
that some substitutions may disrupt a protein's function in ways
other than affecting its stability.
Therefore, for proteins with a high fraction of functional residues,
our theory provides only an upper bound
on the fraction of functional proteins.  However,
our theory's remarkable success for both the
subtilisin and the TEM1 mutant libraries suggests
that this assumption is also valid.

Our theory provides a quantitative rationale 
for earlier work with lattice proteins on the organization of
functional proteins in sequence space.  Bornberg-Bauer and 
Chan~\cite{Bornberg-Bauer1999} proposed that proteins are located
in superfunnels in sequence space with the most stable
sequence having the most neutral neighbors; others have reported
that folded proteins surround highly stable prototype sequences
in sequence space~\cite{Taverna2002,Wingreen2004,Xia2004b}, and
Shakhnovich and coworkers~\cite{Broglia1999} showed 
that proteins with a large energy gap
between the lowest and second lowest energy conformations are
stabilized against mutations.  We
provide a clear explanation: more stable proteins are able to tolerate 
more of the possible mutations before unfolding, 
and so a higher fraction of their neighboring sequences fold.

In addition to these stability-based effects, different 
protein structures have different inherent designabilities,
with more sequences folding into some structures than
others~\cite{Li1996,Wolynes1996,England2003a}.  
Proteins with more designable structures might be 
expected to show a higher average neutrality
since their structures occupy a larger fraction of sequence space.
The average neutrality \nuaa\ therefore provides a 
quantitative measure of designability that can be estimated with
current computational techniques.

Our work suggests a more nuanced approach to experimentally
analyzing protein neutralities than has been applied in the past.
Loeb and coworkers~\cite{Guo2004} have performed a careful 
analysis of the neutralities of several proteins or
regions of proteins under the assumption of a strict exponential
decline in $m$-neutrality.  However, our work suggests that a
protein's $m$-neutrality can deviate from a strict exponential for
the first few substitutions if the protein has a large
amount of extra stability, as we show for the M182T variant
of TEM1.  Experimental mutagenesis studies suggest that during natural
evolution, proteins accumulate mildly destabilizing mutations that
are counterbalanced by stabilizing mutations~\cite{Serrano1993}.
We suggest that it is also important to examine whether some 
natural proteins have systematically accumulated
stabilizing mutations in order to provide them with additional
robustness~\cite{Wilke2002} to amino acid substitutions.

Our work also has applications in protein
engineering.  Directed evolution involves 
screening libraries of mutant proteins for new or improved
functions~\cite{Arnold1998}.  Each round of
directed evolution typically introduces only one or two 
amino acid substitutions
because the rapid decline in $m$-neutrality means 
that higher mutation rates will yield libraries of
mostly unfolded proteins. 
Our work suggests that using 
highly stable parents for directed evolution should increase the
fraction of folded mutants at a given level of substitutions.  It also
provides a method for predicting which structures will
better tolerate large numbers of substitutions. 

\small 
\subsection*{Acknowledgements}
We thank Brian Shoichet for providing us with genes for the 
TEM1 $\beta$-lactamase variants, and Titus Brown for programming assistance.
We thank Michelle Meyer and Eric Zollars 
for helpful
advice and discussions, and two anonymous reviewers for their helpful 
comments.  J.D.B. is supported by a Howard Hughes 
Medical Institute predoctoral fellowship.
D.A.D. is supported by the National Institutes of Health, National Research
Service Award 5 T32 MH19138 from the National Institute of Mental Health.
C.A. and C.O.W. were supported in part by the National Science Foundation
under grant DEB-9981387.

\newpage

\begin{figure*}
\begin{psfrags}
\psfrag{0}[c][c]{0}
\psfrag{3}[c][c]{3}
\psfrag{6}[c][c]{6}
\psfrag{9}[c][c]{9}
\psfrag{number of clones}[c][c]{\large number of clones}
\psfrag{number of mutations}[c][c]{\large number of mutations}
\centerline{\includegraphics[width=4.0in]{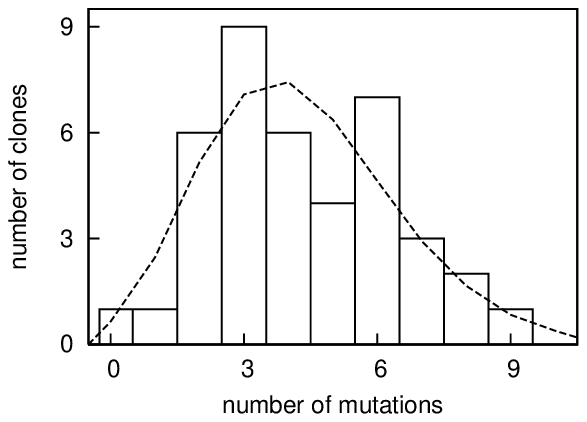}}
\end{psfrags}
\caption{The observed distribution of mutations among the $40$
sequenced unselected TEM1 $\beta$-lactamase
clones from the round-five mutant libraries
(bars) agrees with the theoretical
predictions (lines) for the distribution of mutations in an error--prone
PCR library.  The theoretical predictions are made
with the equation $f\left(\mnt\right)$ described in the text, with
an average number of nucleotide mutations of $\mntavg = 4.3$
for the 570 bp sequenced region of the 861 bp gene.
A chi-square test demonstrated that the
observed distribution is consistent
with the theoretical predictions, with a $P$-value $> 0.9$ that
a difference at least this large between the observed and
predicted values would occur by chance.
}
\end{figure*}

\begin{table*}
\center{\bf Mutations in the round five TEM1 libraries} 
\medskip

\centerline{\small{\begin{tabular}{|l|l|l|} \hline
{\bf Clone} & {\bf TEM1 WT} & {\bf TEM1 M182T} \\
\hline
1 & G43A (A15T), T64C (F22L),  & G122T (R41L), A158T (K53M),  \\
   & C68A (A23D), A90G, G154A (G52S),  & A185C (E62A), G198A (M66I),  \\
   & A375C, T407C (L136P), A426G,  & C204T, T246C, T408C \\
   & T479G (L160R) &  \\
\hline
2 & T52C (C18R), A71C (H24P),  & A94G (K32E), T161A (I54N),  \\
   & A149G (N50S), T200A (M67K),  & A195T, G385A (D129N), A430T (STOP),  \\
   & A319C (T107P), T332A (L111H),  & A467G (H156R), T470A (V157E) \\
   & A388T (N130Y), T393C &  \\
\hline
3 & C28A (L10I), delT30, T38C (F13S),  & T49C (F17L), T50C (F17S),  \\
   & T50C (F17S), A123T, A151G (S51G),  & A120G, delG128, C249T,  \\
   & T190C (F64L), G327C (K109N) & T278C (I93T) \\
\hline
4 & T42C, T200G (M67R), T384A (S128R),  & T33A, T130C (Y44H), A264G,  \\
   & A423G, G459T (M153I), A503C (N168T),  & G268A (G90S), C466T (H156Y),  \\
   & delT504 & C545T (A182V) \\
\hline
5 & A77G (E26G), G100A (A34T),  & A90G, T223A (C75S), C241T (R81C),  \\
   & A149G (N50S), C177T, T236A (STOP),  & A263G (Q88R), T339A (D113E),  \\
   & A418T (I140F) & T410A (L137Q) \\
\hline
6 & T171G (S57R), A186G, T317C (V106A),  & T161C (I54T), T243C, A364G (S122G),  \\
   & T465C, A518G (N173S), A549G & C427T (P143S), A550G (M184V) \\
\hline
7 & C27T, T52A (C18S), G67A (A23T),  & T69A, T174C, A195G,  \\
   & T216C, T374C (I125T), A491G (E164G) & G488A (STOP), A559G (T187A) \\
\hline
8 & T39C, T40C (F14L), T51A (F17L),  & T41C (F14S), T302A (V101D),  \\
   & T361A (C121S), A472G (T158A),  & A467G (H156R), G476A (R159H),  \\
   & G476A (R159H) & A482G (D161G) \\
\hline
9 & A158T (K53M), A186G, G306A,  & T144C, C276T, A281G (H94R),  \\
   & T384G (S128R), T539G (M180R) & A513G (I171M) \\
\hline
10 & T41A (F14Y), T86A (V29E),  & A148T (N50Y), C153T, A199G (M67V),  \\
   & A464G (D155G), A570G & C378T \\
\hline
11 & delA7, T245A (V82D), T332C (L111P) & C70T (H24Y), T245G (V82G),  \\
   &  & T286A (S96T), A357G \\
\hline
12 & A95G (K32R), A308T (Y103F),  & T33C, C147T, A348G,  \\
   & G525A & T404C (L135S) \\
\hline
13 & T447G (F149L), A452G (H151R),  & A95G (K32R), T225A (STOP),  \\
   & A513G (I171M) & G383A (S128N), C416T (T139M) \\
\hline
14 & A321G, A524G (E175G), A538T (M180L) & A258C (Q86H), C267T, A315T \\
\hline
15 & T112C, A167G (E56G), G566A (R189H) & T173A (F58Y), T393C, C539T (T180M) \\
\hline
16 & T112C, A158G (K53R) & G340A (G114S), C436T, A454G (N152D) \\
\hline
17 & C101G (A34G), T384A (S128R) & A157T (STOP), T403G (L135V),  \\
   &  & C440T (T147I) \\
\hline
18 & A158G (K53R), T233A (V78E) & delT42, T380A (M127K) \\
\hline
19 & C28T (L10F), A516C & G168T (E56D), C187T (R63C) \\
\hline
20 &  & A352T (STOP) \\
\hline
\end{tabular}}}
\caption{\label{tab:mutations}
The mutations found in the $20$ unselected clones from the round
five wildtype and M182T TEM1 $\beta$-lactamase libraries.
For nonsynonymous mutations, the amino acid change is shown in 
parentheses.}
\end{table*}

\end{document}